\documentclass[5p]{elsarticle}

\usepackage{lineno}

\usepackage{amssymb}
\usepackage{amsmath}
\usepackage{tikz}
\usetikzlibrary{shapes,arrows}
\usepackage{xcolor}
\journal{SoftwareX}
\usepackage{enumitem}
\usepackage{setspace}
\usepackage{parskip}
\usepackage{multicol}
\usepackage{multirow}
\usepackage[hidelinks,colorlinks]{hyperref}
\usepackage{listings}

\biboptions{sort&compress}

\newcommand{\email}[1]{\href{mailto:#1}{#1}}
\newcommand{\sub}[1]{\ensuremath{_{\text{#1}}}}

\newcommand{\rmd}[1]{\ensuremath{\text{d}}}
\newcommand{\longleftcircarrow}{\ensuremath{\bullet\hspace{-0.5em}-\hspace{-0.5em}-\hspace{-0.5em}-}}
\definecolor{DarkRed}{rgb}{0.7,0,0}
\definecolor{DarkGreen}{rgb}{0,0.7,0}
\definecolor{DarkBlue}{rgb}{0,0,0.7}

\begin{document}

\begin{frontmatter}

\title{JDFTx: software for joint density-functional theory}

\author[1]{Ravishankar Sundararaman\corref{corr}}\ead{sundar@rpi.edu}
\author[2]{Kendra Letchworth-Weaver}
\author[3]{Kathleen A. Schwarz}
\author[4]{\\Deniz Gunceler}
\author[4]{Yalcin Ozhabes}
\author[4]{T.A. Arias}

\address[1]{Department of Materials Science and Engineering, Rensselaer Polytechnic Institute, Troy, NY, 12180}
\address[2]{Center for Nanoscale Materials, Argonne National Laboratory, Lemont, IL 60439}
\address[3]{National Institute of Standards and Technology, Material Measurement Laboratory, Gaithersburg, MD, 20899}
\address[4]{Department of Physics, Cornell University, Ithaca, NY 14853}
\cortext[corr]{Corresponding author}

\makeatletter{}\begin{abstract}
Density-functional theory (DFT) has revolutionized computational prediction of
atomic-scale properties from first principles in physics, chemistry and materials science.
Continuing development of new methods is necessary for accurate predictions
of new classes of materials and properties, and for connecting to nano-
and mesoscale properties using coarse-grained theories.
JDFTx is a fully-featured open-source electronic DFT software designed specifically
to facilitate rapid development of new theories, models and algorithms.
Using an algebraic formulation as an abstraction layer, compact C++11
code automatically performs well on diverse hardware including GPUs (Graphics Processing Units).
This code hosts the development of joint density-functional theory (JDFT)
that combines electronic DFT with classical DFT and continuum models of liquids
for first-principles calculations of solvated and electrochemical systems.
In addition, the modular nature of the code makes it easy to extend and
interface with, facilitating the development of multi-scale toolkits that
connect to \emph{ab initio} calculations, e.g. photo-excited carrier dynamics
combining electron and phonon calculations with electromagnetic simulations.
\end{abstract}


\end{frontmatter}

\makeatletter{}\section{Motivation and significance}
\label{sec:Motivation}

Density functional theory (DFT) enables computational prediction of
material properties and chemical reactions starting from
a quantum mechanical description of the electrons.
DFT codes are now widely used to understand and design new materials
from first principles through the prediction of electronic properties, 
structures and dynamics of molecules, solids and surfaces.
Such studies commonly employ proprietary software such as GAUSSIAN \cite{GAUSSIAN}
and VASP \cite{VASP} as well as open-source software such as
Quantum Espresso \cite{QE}, ABINIT \cite{ABINIT} and Qbox \cite{Qbox},
to name just a few.\footnote{Certain commercial software codes are identified in this paper to foster understanding.
Such identification does not imply recommendation or endorsement by the National Institute of Standards and Technology,
nor does it imply that the materials or equipment identified are necessarily the best available for the purpose.}

However, DFT offers limited accuracy for certain classes of materials and
properties \cite{DFTinaccuracy}, and is extremely computationally expensive
for amorphous materials, liquids and nanostructures \cite{AmorphousInterface}.
The study of new systems, as soon as they become technologically and scientifically relevant,
requires continual development of new methods to improve the accuracy of DFT
and incorporate it into multi-scale theories to access higher length scales.
Developing and testing such methods within production codes is extremely challenging and time consuming.

Systems involving liquids, such as electrochemical interfaces or solvated biomolecules, are particularly
challenging for DFT calculations, requiring thermodynamic sampling of several thousands
of atomic configurations in \emph{ab initio} molecular dynamics (AIMD) simulations \cite{CPMD}.
Joint density-functional theory (JDFT) was proposed as a theoretical framework
to address this issue by combining electronic DFT with classical DFT of liquids \cite{CDFT-review}
to directly compute equilibrium properties of quantum-mechanically
described solutes in diverse solvent environments \cite{JDFT}.
Bringing this method to fruition required the simultaneous development of
physical models (free energy functionals) of liquids and their interaction
with electrons, algorithms to perform variational free-energy minimization
and code that tightly and efficiently coupled these new models with electronic DFT.
We began the open-source software project JDFTx in 2012 to facilitate
this combined model, algorithm and code development effort.

\makeatletter{}\lstset{language=C++,basicstyle={\linespread{0.8}\small\ttfamily},basewidth={.5em,.5em},
	keywordstyle=\color{DarkRed},commentstyle=\color{DarkBlue}}
\begin{lstlisting}[float=*, xleftmargin=0.1\textwidth, label=lst:LinearPCM, caption={
	JDFTx code for a simplified solvation model with local dielectric screening.
	The extended DFT++ algebraic formulation \cite{AlgebraicDFT,RigidCDFT}
	expresses the physical model in almost the same language as the defining
	equations (\ref{eq:Poisson}) and (\ref{eq:Adiel}), while parallel implementations
	of the operators make this code, \emph{written only once}, run with 
	multiple CPU threads (using pthreads) or on NVIDIA GPUs (using CUDA).
}]
class LinearPCM : public LinearSolvable<ScalarFieldTilde>
{  ScalarField epsilon; //inhomogeneous dielectric
public:
   virtual ScalarFieldTilde hessian(const ScalarFieldTilde& phiTilde) const
   {  return (-1./(4*M_PI)) * divergence(J(epsilon * I(gradient(phiTilde))));
   }
   double getAdiel(const ScalarFieldTilde& rhoTilde) const
   {  solve(rhoTilde); //solve hessian(state) = rhoTilde
      return 0.5 * dot(state - coulomb(rhoTilde), O(rhoTilde));
   }
};
\end{lstlisting}

This article introduces JDFTx as a general-purpose user-friendly DFT software
that offers a full feature set, yet is simultaneously developer-friendly to
enable rapid prototyping of new electronic-structure and related methods.
Section~\ref{sec:Description} presents the overall design of JDFTx using
the algebraic formulation of DFT \cite{AlgebraicDFT,RigidCDFT} to
separate implementation into physics, algorithm and hardware layers,
enabling rapid development of high-performance code that is easy to use.
It also outlines commonly used features of the code, some of which are
illustrated in more detail with examples in section~\ref{sec:Examples}.
Finally, section~\ref{sec:Impact} highlights new methods that have
already been developed using JDFTx, including a hierarchy of JDFT models
for the electronic structure of solvated systems and a toolkit for
photo-excited carrier dynamics with \emph{ab initio} electron
and phonon properties, as well as key applications of these methods.

\section{Software description}
\label{sec:Description}

The core functionality of any electronic DFT software includes the calculation
of ground-state electron densities, energies and forces within the Kohn-Sham
DFT formalism \cite{KS-DFT}, given a list of atoms and their positions.
This facilitates prediction of structure and dynamics of materials,
evaluation of reaction pathways and chemical kinetics, as well as
determination of phase equilibria and stability.

Due to the nature of quantum-mechanical simulations of matter, DFT calculations become
increasingly expensive with the number of atoms and electrons involved;
computational complexity ranges from $O(N^3)$ (with a smaller prefactor)
to $O(N)$ (with a much larger prefactor), depending on the implementation.
A brute force approach to nano and mesoscale systems with several
thousands to millions of atoms is therefore not practical;
it is instead logical to develop multi-scale theories for the properties
of interest while still incorporating DFT electronic structure where appropriate.

JDFTx is an open-source DFT software designed specifically with the goals of coupling 
electronic DFT with coarse-grained theories to bridge atomic and system length scales,
and of facilitating the rapid development of new classes of such combined theories.
It implements electronic DFT in the plane-wave basis, which is best suited for periodic
systems such as solids and solid surfaces, but is also applicable to molecular systems.
A key functionality of JDFTx beyond standard electronic DFT codes is
the modeling of liquids using classical DFT \cite{RigidCDFT}, and
JDFT calculations of electronic structure in liquid environments by
combining electronic DFT with classical DFT or simpler solvation models.
Section~\ref{sec:Architecture} presents a birds-eye view of the code architecture
along with a code example to illustrate the ease of developing new features,
after which section~\ref{sec:Functionalities} outlines the key features of the code.

\begin{figure*}
\includegraphics[width=\textwidth]{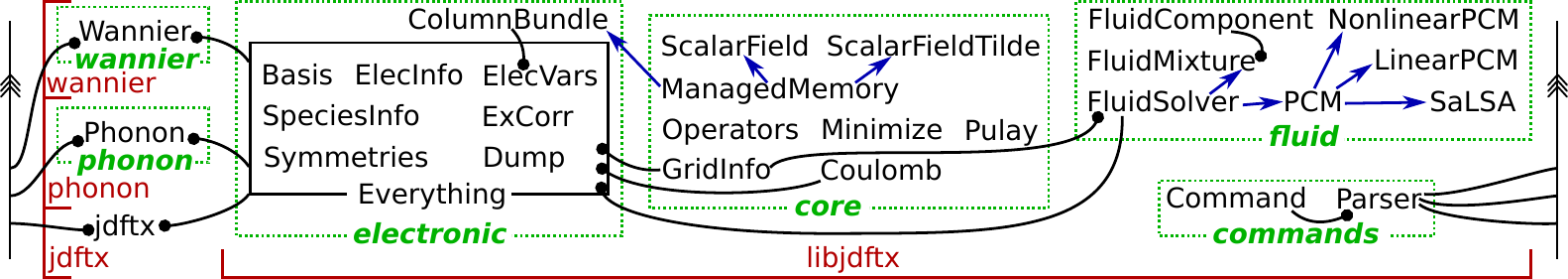}
\caption{Organization of the JDFTx code:
key classes/files shown in \textcolor{black}{black},
organized into sub-directories shown in \textbf{\textcolor{DarkGreen}{green}}
and compilation targets (executables/libraries) shown in \textcolor{DarkRed}{red}.
$A \longleftcircarrow B$ indicates $A$ contains $B$,
while $A \textcolor{DarkBlue}{\longleftarrow} B$ indicates $A$ inherits from $B$.
Lightweight executables jdftx, phonon and wannier link to a dynamic link library, libjdftx,
which contains most of the functionality and can be readily used from third-party codes.
With CUDA available, compilation of the \emph{same} code additionally results in
jdftx\_gpu, phonon\_gpu etc. that run almost entirely on compute-capable GPUs.
Only a small fraction of files and connections are shown here;
see the API documentation at \url{http://jdftx.org} for more details.
\label{fig:Architecture}}
\end{figure*}

\subsection{Software Architecture}
\label{sec:Architecture}

JDFTx achieves its goal of code simplicity and extensibility by using
the `DFT++' algebraic formulation of electronic DFT \cite{AlgebraicDFT}
and its generalization to classical DFT and JDFT \cite{RigidCDFT}.
This algebraic formulation cleanly separates the code into physics,
algorithm and computational layers.
Theories and algorithms are expressed concisely at a high-level of abstraction
in the top layers of the code, whereas performance optimizations and support
for specialized hardware such as GPUs are handled in the lower layers.

We illustrate this clean separation with an example of a simplified solvation model defined by
\begin{align}
-4\pi\rho(\vec{r}) &= \nabla\cdot(\epsilon(\vec{r})\nabla\phi(\vec{r})),\quad\text{and} \label{eq:Poisson}\\
A\sub{diel} &= \frac{1}{2}\int\rmd\vec{r} \left[ \phi(\vec{r}) - \hat{K} \rho(\vec{r}) \right] \rho(\vec{r}). \label{eq:Adiel}
\end{align}
Here, the liquid is treated as an inhomogeneous dielectric $\epsilon(\vec{r})$
which interacts with the charge density of the electronic system, $\rho(\vec{r})$.
The net electrostatic potential $\phi(\vec{r})$ satisfies the modified
Poisson equation (\ref{eq:Poisson}), and the electrostatic solvation energy
$A\sub{diel}$ is the difference between the dielectric-screened and unscreened
electrostatic self energies of $\rho(\vec{r})$ (\ref{eq:Adiel}),
where $\hat{K}$ is the unscreened Coulomb operator.
This is the essence of most solvation models used with
DFT \cite{PCM-SMD, PCM-SCCS, NonlinearPCM, VASPsol};
we have only skipped the determination of $\epsilon(\vec{r})$
from atom positions or electron densities, and additional
non-electrostatic correction terms in $A\sub{diel}$ for brevity.
Regardless, solving the Poisson equation above is the most
complex and time-consuming operation in these solvation models.

\newcommand{\entity}[1]{\emph{#1}}

Listing~\ref{lst:LinearPCM} shows the implementation of this model in JDFTx.
Class \entity{LinearPCM} derives from a templated abstract base class \entity{LinearSolvable},
which implements the Conjugate Gradients (CG) algorithm \cite{CG} on arbitrary vector spaces,
instantiated in this case for scalar fields in reciprocal space, \entity{ScalarFieldTilde}.
The equation to be solved is defined by the virtual function \entity{hessian}, whose
one-line implementation can be recognized as (\ref{eq:Poisson}) at a moment's glance.
Note that the \entity{gradient} ($\nabla$) and \entity{divergence} ($\nabla\cdot$) operators
apply in reciprocal space (where they are diagonal), while the operators
\entity{I} and \entity{J} Fourier transform from reciprocal to real space and vice versa
[\cite{AlgebraicDFT}] (using Fast Fourier Transforms).
The function \entity{getAdiel} first calls function \entity{solve} from base class LinearSolvable
to solve (\ref{eq:Poisson}) for $\phi(\vec{r})$ stored in a member variable \entity{state}
of the base class, and the second line evaluates $A\sub{diel}$ using (\ref{eq:Adiel}).
The integral is evaluated as a dot product with the overlap operator \entity{O},
and \entity{coulomb} implements the unscreened Coulomb operator $\hat{K}$,
both of which are diagonal in the plane-wave basis [\cite{AlgebraicDFT}].

The algebraic formulation enables Listing~\ref{lst:LinearPCM} to resemble
(\ref{eq:Poisson}) and (\ref{eq:Adiel}) as closely as possible,
streamlining the translation of physics into code.
In addition, the linear algebra representation allows the implementation of all involved operators
(eg. \entity{I},\entity{J}, *, \entity{gradient} etc.) to be optimized under the hood for different hardware.
In particular, all operators in JDFTx are already implemented using
both pthreads for multi-core CPUs (Central Processing Units) and CUDA for NVIDIA GPUs.
This division of labor allows \emph{all} physics code in JDFTx to be implemented only once,
yet run natively on all supported hardware configurations.  Consequently, JDFTx could compute
entirely on GPUs since its inception in 2012 (a first for plane-wave DFT).

Additionally, the use of C++11 features such as rvalue references
and smart pointers in the implementation of data structures
(eg. ScalarField) and their operators helps minimize memory overhead.
For example, in hessian() in Listing~\ref{lst:LinearPCM},
\entity{I()}, \entity{epsilon *} and \entity{J()} automatically operate in-place because
their inputs are temporary return values of other functions.

Fig.~\ref{fig:Architecture} shows the organization of the key classes and files of the
JDFTx codebase, containing 275 source files with approximately 60000 lines of code; this fairly compact
implementation for a DFT code is possible because of the algebraic framework discussed above.
Most of the functionality is compiled into a dynamic link library \entity{libjdftx},
which is accessed through light-weight executables \entity{jdftx} for most calculations,
\entity{phonon} for phonon dispersion and electron-phonon interaction calculations,
and \entity{wannier} for generation of maximally-localized Wannier
functions \cite{MLWF} and \emph{ab initio} tight binding models.
This organization makes the procedure straightforward for other DFT codes
to leverage JDFTx features, especially JDFT and related
solvation models, by linking directly to libjdftx.

The \entity{core} code (Fig.~\ref{fig:Architecture}) implements the basic
data structures, operators and algorithms used by the rest of JDFTx.
\entity{ManagedMemory} handles data transfers between CPUs and GPUs
(automatically, when necessary), and is the base class for
\entity{ScalarField}s and electronic wavefunctions in \entity{ColumnBundle}.
\entity{Minimize} and \entity{Pulay} provide algorithms for variational minimization,
including linear and nonlinear CG \cite{CG}, \entity{L-BFGS} \cite{LBFGS})
and self-consistency by Pulay mixing \cite{PulayMixing}.
\entity{GridInfo} describes the plane-wave grid and Fourier transforms,
while \entity{Coulomb} implements the Coulomb kernel in various dimensions \cite{TruncatedEXX}.

The \entity{fluid} code is tied together by the abstract base class \entity{FluidSolver},
which is implemented by \entity{FluidMixure} containing several \entity{FluidComponent}s
to provide the classical DFTs \cite{RigidCDFT,PolarizableCDFT} used for JDFT; see Ref.~\cite{RigidCDFT}
for more details on this framework to implement atomically-detailed classical DFT for molecular
fluids in terms of \entity{IdealGas} representations and \entity{Fex} (excess) functionals.
Alternately, a hierarchy of solvation models with linear (LinearPCM \cite{PCM-Kendra}),
nonlinear (NonlinearPCM \cite{NonlinearPCM}) and even nonlocal response (SaLSA \cite{SaLSA})
derive from base class \entity{PCM}.

\begin{table*}
\centering
\makeatletter{}\newenvironment{myitemize}
{ \begin{itemize}[leftmargin=.2in]
    \setlength{\itemsep}{0pt}}
{ \end{itemize}                  } 

\framebox{\noindent\begin{minipage}{6.5in}\small
\begin{minipage}{0.49\columnwidth}
{\bf Electronic}
\begin{myitemize}
\item Exchange-correlation: semilocal, meta-GGA,\\EXX-hybrids, DFT+$U$, DFT-D2, LibXC
\item Pseudopotentials: norm-conserving and ultrasoft
\item Noncollinear magnetism / spin-orbit coupling
\item Algorithms: varitional minimization, SCF
\item Grand canonical (fixed potential) for electrochemistry
\item Truncated Coulomb for 0D, 1D, 2D or 3D periodicity
\item Custom external potentials, electric fields
\item Charged-defect corrections: bulk and interfacial
\item Ion/lattice optimization with constraints
\item \emph{Ab initio} molecular dynamics
\item Vibrational modes, phonons and free energies
\end{myitemize}
\end{minipage}
\begin{minipage}{0.49\columnwidth}
{\bf Fluid}
\begin{myitemize}
\item Linear solvation: GLSSA13, SCCS, CANDLE
\item Nonlinear solvation: GLSSA13
\item Nonlocal solvation: SALSA
\item JDFT with classical DFT fluids
\end{myitemize}
{\bf Outputs (selected)}
\begin{myitemize}
\item DOS, optical matrix elements, polarizability etc.
\item Wannier functions and \emph{ab initio} tight-binding
\item Electron-electron and electron-phonon scattering
\end{myitemize}
{\bf Interfaces}
\begin{myitemize}
\item Solvated QMC with CASINO
\item Atomistic Simulation Environment (NEB, MD etc.)
\item Visualization: VESTA, XCrySDen, PyMOL
\end{myitemize}
\end{minipage}
\end{minipage}}

\caption{Selected features of JDFTx. See index of input file
commands on \url{http://jdftx.org} for a complete list.}
\label{tab:Features}
\end{table*}

The \entity{electronic} code contains the standard Kohn-Sham electronic DFT implementation
(with class names derived from the original implementation of DFT++ \cite{AlgebraicDFT}).
Here, \entity{ElecInfo} and \entity{ElecVars} contain electronic occupations, wavefunctions,
density, potential etc. and the functions relating them.
\entity{SpeciesInfo} handles electron-ion interactions (pseudopotentials),
\entity{ExCorr} implements exchange-correlation functionals, and
\entity{Dump} handles outputs and post-processing.
All this functionality (including fluids) is tied together
into the container class \entity{Everything} for convenience.

Finally, the \entity{commands} code (Fig.~\ref{fig:Architecture}) provides an
object-oriented interface to define commands and parse input files.
All executables use this module to provide a consistent input file syntax
that supports environment variable substitution and modular input files;
\entity{phonon} and \entity{wannier} support all jdftx commands in addition to those
specific to phonon dispersion and Wannier function calculations respectively.
Complete description of the classes, files and connections between them
is available in the API documentation generated automatically by
Doxygen \cite{Doxygen} on the JDFTx website, \url{http://jdftx.org}.

\subsection{Software Functionalities}
\label{sec:Functionalities}

Table~\ref{tab:Features} lists a selection of features available in JDFTx.
It supports the full range of functionality found in all major electronic DFT software.
It has built-in support for several semilocal \cite{LDA,PBE,PBEsol}, meta-GGA \cite{revTPSS}
and EXX-hybrid \cite{PBE0,HSE06,HSE12} exchange-correlation functions,
with additional functionals available through LibXC \cite{LibXC}.
DFT+$U$ \cite{DFTplusU} and DFT-D2 \cite{Dispersion-Grimme} pair potential dispersion
corrections allow for handling localized electrons and van der Waals interactions respectively.
JDFTx supports norm-conserving and ultrasoft pseudopotentials in the FHI, UPF and USPP
formats, which allows easy interoperability with QE \cite{QE} and ABINIT \cite{ABINIT}.
It automatically installs two well-tested, open-source pseudopotential libraries,
GBRV (ultrasoft) \cite{GBRV} and SG15 (norm-conserving) \cite{SG15},
enabling out-of-the-box calculations for most elements.
JDFTx supports interactions with custom external potentials and fields,
and allows accurate calculations of systems of any dimensionality:
molecules (0D), wires (1D), slabs / 2D materials and bulk (3D),
using truncated Coulomb interactions \cite{TruncatedEXX}.

Importantly, JDFTx implements two distinct classes of algorithms for
electronic DFT, variational minimization of the (analytically-continued)
total energy \cite{AnalyticContinuedDFT, AuxiliaryHamiltonian}
and the self-consistent field (SCF) iteration method \cite{DIIS}.
Variational minimization is stable and guaranteed to converge, while SCF
(the default method available in all DFT codes) is less stable
in general, but faster when it converges well.
JDFTx also uniquely implements grand-canonical DFT \cite{GC-DFT}, where
electron number adjusts automatically at a fixed electron chemical potential,
which correctly describes the behavior of electrochemical systems.
SCF convergence can be problematic in this mode, and for many solvated systems
in general; variational minimization is therefore essential to have as an alternative.

JDFTx specializes in electronic structure calculations incorporating
a continuum description of the environment, with a the range of
techniques available for including fluids and solvation effects.
Full JDFT calculations include a detailed classical DFT description
of the liquid that captures atomic-scale structure in a number
of solvents \cite{BondedVoids, RigidCDFT, PolarizableCDFT}.
JDFTx also includes a hierarchy of solvation models for computationally
inexpensive treatment of solvation effects, ranging from the
nonlocal solvation model SaLSA \cite{SaLSA} which captures
atomic-scale liquid structure at a linear-response level,
through models that capture nonlinear dielectric and ionic
response \cite{NonlinearPCM}, to the simplest linear-response models \cite{PCM-Kendra}.
These linear response models include GLSSA13 \cite{NonlinearPCM}
(later ported to VASP as VASPsol \cite{VASPsol}),
SCCS \cite{PCM-SCCS} (the solvation model available in QE)
and CANDLE \cite{CANDLE}.
Table~\ref{tab:SolvComparison} compares the accuracy of these solvation models
for a standard set of neutral solutes, cations and anions in water,
and shows that CANDLE achieves the best accuracy for aqueous solvation
of charged species by explicitly capturing charge asymmetry in solvation.

\begin{table}
\centering
\setlength\tabcolsep{2pt}
\begin{tabular}{|l|ccc|c|}
\hline
\multirow{2}{*}{Model} & \multicolumn{4}{c|}{MAE [kcal/mol]} \\\cline{2-5}
& Neutral & Cations & Anions & All \\
\hline
Nonlocal SaLSA \cite{SaLSA}             & 1.36 & 3.20 & 19.7 & 4.55 \\
\hline
Nonlinear GLSSA13 \cite{NonlinearPCM}   & 1.28 & 16.1 & 27.0 & 7.55 \\
\hline
Linear GLSSA13 \cite{NonlinearPCM}      & \multirow{2}{*}{1.27} & \multirow{2}{*}{2.10} & \multirow{2}{*}{15.1} & \multirow{2}{*}{3.59} \\
\qquad = VASPsol \cite{VASPsol}         &      &      &      &      \\
SCCS neutral fit 1 \cite{PCM-SCCS}      & 1.20 & 2.55 & 17.4 & 3.97 \\
SCCS neutral fit 2 \cite{PCM-SCCS}      & 1.28 & 2.66 & 16.9 & 3.97 \\
SCCS cation fit \cite{PCM-SCCS-charged} &  --  & 2.26 &  --  &  --  \\
SCCS anion fit \cite{PCM-SCCS-charged}  &  --  &  --  & 5.54 &  --  \\
CANDLE \cite{CANDLE}                    & 1.27 & 2.62 & 3.46 & 1.81 \\
\hline
\end{tabular}
\caption{Comparison of mean absolute errors (MAE) of solvation energies predicted by various
solvation models for an identical set of 240 neutral solutes, 51 cations and 55 anions in water.
See Ref.~\citenum{PCM-SCCS-charged} for a detailed specification of individual solutes.
CANDLE provides the best accuracy for neutrals, cations and anions within a single parametrization
because it explicitly accounts for charge asymmetry in solvation \cite{CANDLE}.
\label{tab:SolvComparison}}
\end{table}

JDFTx can export a wide-range of electronic structure and liquid properties including
charge/site densities, potentials, density of states, vibrational/phonon modes
and free energies, optical and electron-phonon matrix elements etc.
(See \url{http://jdftx.org/CommandDump.html} for a full list.)
Executable \entity{wannier} can generate maximally-localized Wannier functions for
either separated or entangled bands \cite{MLWF,MLWFmetal} and transform
Hamiltonians and matrix elements into an \emph{ab initio} tight-binding model.
JDFTx is also interfaced with several commonly-used visualization software packages,
and with the Atomistic Simulation Environment \cite{ASE} for features
including Nudged Elastic Band (NEB) \cite{NEB} barrier calculations
and alternate molecular dynamics methods.
It provides solvation functionality to other electronic structure software through
interfaces, such as for Quantum Monte-Carlo (QMC) simulations in CASINO \cite{CASINO}.
 
\makeatletter{}\section{Illustrative Example}
\label{sec:Examples}

JDFTx is used exactly like most plane-wave DFT software, with an
input file that describes the atomic geometry, pseudopotentials,
functionals, and other similar options alongside the operations to be performed.
For example, Listing~\ref{lst:InputFile} shows an
input file for a formate ion on a Pt(111) surface,
modeled as a 3 layer slab in a 2$\times$2 supercell,
with the slab normal along the third lattice vector.
The first section selects the GBRV pseudopotentials \cite{GBRV} for all 
elements (using the wildcard `\$ID') and corresponding kinetic energy
cutoffs (all energies in Hartrees ($E_h$)) for wavefunctions and charge densities.
The second section specifies the lattice geometry,
slab-mode Coulomb truncation to isolate periodic images 
along $z$, ionic positions in lattice coordinates,
and requests 10 iterations of ionic geometry optimization.
The final 0 on the Pt atom positions constrains their positions
completely, the 1 allows the rest to move, while `Planar~0~0~1'
constrains the C atom to only move in the plane normal to the third lattice vector.

\lstset{language=Python,
	keywordstyle=\color{DarkRed},commentstyle=\color{DarkBlue},
	morekeywords={ion,species,elec,cutoff,lattice,coulomb,interaction,truncation,embed,kpoint,
	folding,smearing,target,mu,ionic,minimize,pcm,variant,fluid,solvent,cation,anion,dump,name}}
\begin{lstlisting}[float, label=lst:InputFile, caption={
JDFTx input file demonstrating several useful features:
fixed-potential calculation of a formate ion on Pt(111) using
the CANDLE solvation model for a 1M aqueous NaF electrolyte,
geometry optimization with Pt positions and $z$ coordinate of C constrained,
and with electron density and fluid bound charge output.
}]
#--- Pseudopotentials ---
ion-species GBRV/$ID_pbe_v1.2.uspp   #GBRV family
ion-species GBRV/$ID_pbe_v1.uspp     #GBRV family
elec-cutoff 20 100         #Ecuts for psi and rho
#--- Geometry ---
lattice Hexagonal 10.53 30.0    #a and c in bohrs
coulomb-interaction Slab 001  #Make z nonperiodic
coulomb-truncation-embed 0 0 0    #Specify center
coords-type Lattice       #fractional coordinates
ion Pt  0.33333 -0.33333 -0.288   0
ion Pt  0.33333 -0.83333 -0.288   0
ion Pt  0.83333 -0.83333 -0.288   0
ion Pt  0.83333 -0.33333 -0.288   0
ion Pt  0.16667 -0.16667 -0.144   0
ion Pt  0.16667 -0.66667 -0.144   0
ion Pt  0.66667 -0.16667 -0.144   0
ion Pt  0.66667 -0.66667 -0.144   0
ion Pt  0.000    0.000    0.000   0
ion Pt  0.000   -0.500    0.000   0
ion Pt  0.500    0.000    0.000   0
ion Pt  0.500   -0.500    0.000   0   #0 => fixed
ion O   0.152   -0.079    0.155   1   #1 =>  free
ion O  -0.152   +0.079    0.155   1
ion C   0.000    0.000    0.190   1  Planar 0 0 1
ion H   0.000    0.000    0.260   1
ionic-minimize nIterations 10  #Optimize geometry
#--- Electronic ---
kpoint-folding 6 6 1       #Gamma-centered k-mesh
elec-smearing Cold 0.01     #Select cold smearing
target-mu -0.160             #Fix echem potential
#--- Fluid ---
fluid LinearPCM         #Class of solvation model
pcm-variant CANDLE   #Specific model within class
fluid-solvent H2O            #Aqueous electrolyte
fluid-cation Na+ 1.           #1 mol/L Na+ cation
fluid-anion   F- 1.           #1 mol/L  F-  anion
#--- Outputs ---
dump Ionic IonicPositions ElecDensity BoundCharge
dump-name test.$VAR      #Output filename pattern
\end{lstlisting}

The third section of Listing~\ref{lst:InputFile} selects Brillouin 
zone sampling, smearing, and a unique feature of JDFTx: grand canonical
DFT \cite{GC-DFT} at a fixed electron chemical potential $\mu = -0.16~E_h$.
The fourth section selects the CANDLE solvation model \cite{CANDLE}
for water with 1M (mol/L) Na+ and F- ions.
The final section requests output every ionic step of ionic positions,
electron density and bound charge density in the fluid,
with files named using the pattern `test.*'.
The organization above is only for illustration; commands may appear in
any order, may be split over multiple input files with `include' statements,
and may invoke environment variable substitution for ease of scripting.
See \url{http://jdftx.org/Commands.html} for more details.

\begin{figure}
\includegraphics[width=\columnwidth]{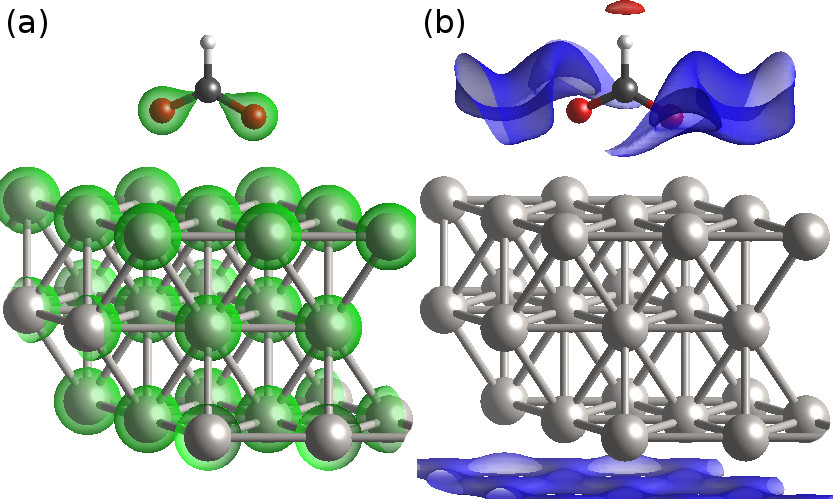}
\caption{Visualization of the structure along with (a) electron density
and (b) fluid bound charge density from the results of running JDFTx
on the input in Listing~\ref{lst:InputFile}, generated using the
createXSF script (distributed with JDFTx) and VESTA \cite{VESTA}.
\label{fig:FormatePt}}
\end{figure}

After saving Listing~\ref{lst:InputFile} to `test.in' (as one possible naming convention),
\lstset{language=C}\begin{lstlisting}
    mpirun -n 4 jdftx -c8 -i test.in -o test.out
\end{lstlisting}
runs the code using 4 MPI processes with 8 threads each,
logging output to `test.out'.  After completion,
\begin{lstlisting}
    createXSF test.out test.xsf n
\end{lstlisting}
creates file `test.xsf' containing the structure and electron
density $n(\vec{r})$, which visualized in VESTA \cite{VESTA}
yields Figure~\ref{fig:FormatePt}(a). The same procedure with
`nbound' instead of `n' yields Figure~\ref{fig:FormatePt}(b)
that shows the bound charge density in the fluid.
As expected, the formate electron density (green) is greatest on
the O atoms, surrounded by a positive (blue) fluid bound charge.
A small negative (red) bound charge appears next to the H,
but at this potential the Pt surface is negatively charged
and mostly surrounded by positive charge in the fluid.
(The raw output from JDFTx uses the opposite electron-is-positive sign
convention rather than the usual convention employed in this discussion.)

\begin{figure}
\includegraphics[width=\columnwidth]{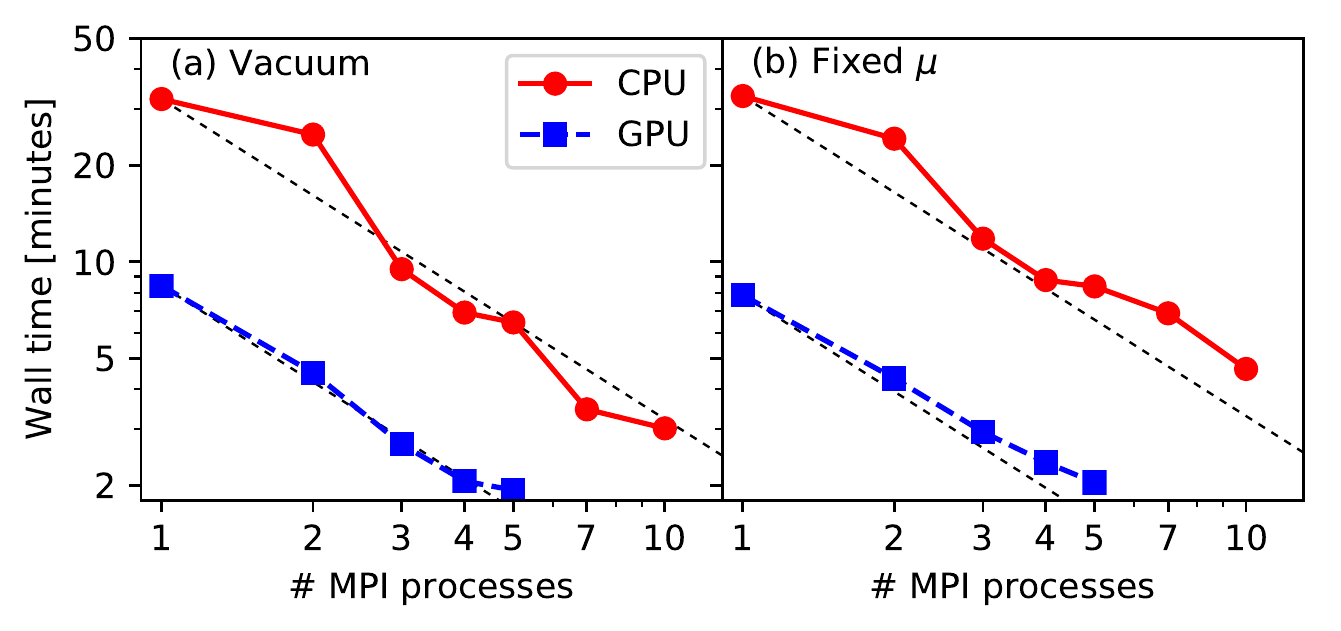}
\caption{Variation of JDFTx run time with number of MPI processes
for (a) the initial vacuum DFT and (b) for the subsequent
grand canonical solvated DFT at the initial geometry
for the calculation specified by Listing~\ref{lst:InputFile}.
Each CPU process runs on one 8-core Intel Xeon E5-2620v4 socket
(two sockets per compute node), while each GPU process
runs on one NVIDIA Tesla K80 GPU (two GPUs per K80 unit).
Dotted lines indicate ideal linear scaling. (The marginal
super-linear speedup for CPU processes in (a) is due to
increased ratio of cache size to memory used per process.)
\label{fig:scaling}}
\end{figure}

Figure~\ref{fig:scaling} illustrates the performance of JDFTx using
the above example for both (a) conventional vacuum DFT calculations
(performed automatically before solvated calculations)
and (b) fixed-potential solvated calculations,
on varying numbers of both CPUs and GPUs.
JDFTx implements MPI parallelization over symmetry-reduced $k$-points
and pthreads / CUDA parallelization over everything else.
It exhibits almost linear MPI scaling when the number of processes
is a factor of this $k$-point count (20 in above example).
Each K80 GPU delivers approximately $3\times$ performance compared to
each 8-core Xeon approximately ($25\times$ compared to each core)
for this problem size.
 
\makeatletter{}\section{Impact}
\label{sec:Impact}

JDFTx simultaneously targets two partially-overlapping research
communities: developers of new DFT methods, models, and algorithms,
and DFT practitioners who take advantage of these new methods.
It emphasizes ease of development as well as use,
realized in practice with highly-modular code that expresses physics in
almost the same language as the theoretical derivations, using the algebraic
formulation of DFT and an efficient C++11 abstraction, as discussed above.
Using this framework, we have rapidly implemented in a few years
a complete feature set comparable to that of proprietary codes
that have been in development for decades,
making JDFTx now usable as a general-purpose DFT software.

JDFTx has facilitated the rapid development of new methods
for a number of applications, most notably for DFT calculations
of solvated and electrochemical systems using JDFT \cite{JDFT} and solvation models.
JDFTx served as the medium for development of liquid free energy functionals
(classical DFT functionals) \cite{BondedVoids,RigidCDFT,PolarizableCDFT} to realize JDFT,
as well as a hierarchy of increasingly accurate solvation models including 
simple linear solvation models \cite{PCM-Kendra,NonlinearPCM,SchwarzCapacitance},
nonlinear models \cite{NonlinearPCM,CavityWDA,PCM-Deniz} and models
incorporating nonlocality: SaLSA \cite{SaLSA} and CANDLE \cite{CANDLE}.
In fact, one of the simplest models developed using JDFTx,
GLSSA13 \cite{NonlinearPCM}, was later ported to the proprietary DFT code
VASP as VASPsol \cite{VASPsol}, the predominant solvation option for that code.
Solvation from JDFTx can be used for quantum Monte Carlo simulations
through an interface with the CASINO code \cite{CASINO,Katie-QMC}.
JDFTx recently enabled the development of grand-canonical DFT algorithms \cite{GC-DFT},
making it possible to realistically simulate electrochemical processes by allowing
number of electrons in the calculation to adjust automatically at fixed potential.
More generally, JDFTx has also been used in electronic structure method development
for exact exchange \cite{TruncatedEXX}, dielectric matrices \cite{Dielectric},
X-ray measurements \cite{TiO2cat},and electron-phonon interactions \cite{PhononAssisted}.

JDFTx makes these cutting-edge methods immediately available in
efficient, easy-to-use code for a wide spectrum of applications.
JDFT and solvation techniques in JDFTx have already been used to
elucidate ion distribution \cite{LiNanoscaleImaging} and dendrite
formation \cite{DendritePrevention1,DendritePrevention2} in Li ion batteries,
surface structures and reaction mechanisms for various 
energy-conversion catalysts \cite{TiO2cat,FormateOxidation,
AnomalousPH,COreductionCu111,IrO2-OER,SurfEnergyCatalysis,
WaterAdHybridPerovskite}, new electrode materials for
supercapacitors \cite{GrapheneCap1,GrapheneCap2,BoropheneCap},
prediction of electrochemical pseudocapacitance \cite{RuO2Cap},
and band alignments at solid-liquid interfaces \cite{BandOffsets,
ElectrostaticPotential}, to take just a few prominent examples.
JDFTx has also facilitated Wannier-function based \emph{ab initio}
tight-binding calculations of photo-excited hot carrier generation
\cite{DirectTransitions,PhononAssisted,MultiPlasmon,HotCarrierReview},
transport \cite{GraphiteHotCarriers,BowtieHotSpots,NESSE}
and ultrafast dynamics \cite{TAparameters,TAanalysis}.
Continuing development of new interfaces and multi-scale methods
using JDFTx will expand the range of applications further.
 
\makeatletter{}\section{Conclusions}

We present JDFTx, a general-purpose open-source plane-wave DFT software,
with a particularly rich feature set for solvated and electrochemical
DFT calculations, and with an emphasis on ease of development and use.
In its first five years, JDFTx has enabled rapid development of
joint density-functional theory (JDFT) and a hierarchy of efficient
and accurate solvation models, which are gradually being implemented
or interfaced with proprietary and other open-source DFT codes.
With these methods, JDFTx has already been widely applied to electrochemical
problems in catalysis, energy conversion and energy storage.

The design of JDFTx allows short, easy-to-read code to perform well
on various hardware architectures, making it ideally suited to
rapidly prototype new methods, and then test and optimize them.
All functionality is exposed as a library, libjdftx,
that other software can link to for direct interfacing.
Future interfaces of JDFTx with other open-source software for
electronic structure calculations, direct computation of experimental observables,
electromagnetic simulations, phase-field methods, and more will drive widespread applications
of multi-scale techniques that start at the electronic scale.

\section*{Acknowledgments}

We acknowledge support (2012-2014) from the Energy Materials Center at Cornell (EMC$^2$),
an Energy Frontier Research Center funded by the U.S. Department of Energy,
Office of Science, Office of Basic Energy Sciences under Award Number DE-SC0001086.
RS acknowledges support (2013-2016) from the Joint Center for Artificial Photosynthesis (JCAP),
a DOE Energy Innovation Hub, supported through the Office of Science
of the U.S. Department of Energy under Award Number DE-SC0004993,
and start-up funding from the Department of Materials Science
and Engineering at Rensselaer Polytechnic Institute.
KLW and KAS acknowledge support from the National Science Foundation
Graduate Research Fellowship.

\bibliographystyle{elsarticle-num} 
\makeatletter{}

\onecolumn

\section*{Required Metadata}

\begin{table}[!h]
\begin{center}
\begin{tabular}{|l|p{6.5cm}|p{6.5cm}|}
\hline
C1 & Current code version                                             & 1.3.1 \\ \hline
C2 & Permanent link to code/repository used for this code version     & \url{https://github.com/shankar1729/jdftx/releases/tag/v1.3.1} \\ \hline
C3 & Legal Code License                                               & GPLv3 \\ \hline
C4 & Code versioning system used                                      & git \\ \hline
C5 & Software code languages, tools, and services used                & C++11, MPI, CUDA \\ \hline
C6 & Compilation requirements, operating environments \& dependencies & GSL, BLAS, LAPACK and FFTW libraries on a POSIX-compliant platform\\ \hline
C7 & Link to developer documentation/manual                           & \url{http://jdftx.org} \\ \hline
C8 & Support email for questions                                      & \email{sundar@rpi.edu} \\ \hline
\end{tabular}
\end{center}
\caption{Code metadata\label{tab:Metadata}}
\end{table}

\begin{table}[!h]
\begin{center}
\begin{tabular}{|l|p{6.5cm}|p{6.5cm}|}
\hline
S1 & Current software version                       & 1.3.1 \\ \hline
S2 & Permanent link to executables of this version  & \url{https://github.com/shankar1729/jdftx/archive/v1.3.1.zip} \\ \hline
S3 & Legal Software License                         & GPLv3 \\ \hline
S4 & Computing platforms/Operating Systems          & POSIX-compliant platform (Linux, Unix, OS X, Windows/Cygwin etc.) \\ \hline
S5 & Installation requirements \& dependencies      & MPI C++11 compiler; GSL, BLAS, LAPACK and FFTW libraries \\ \hline
S6 & Link to user manual                            & \url{http://jdftx.org} \\ \hline
S7 & Support email for questions                    & \email{sundar@rpi.edu} \\ \hline
\end{tabular}
\end{center}
\caption{Software metadata\label{tab:SoftMetadata}}
\end{table}

\end{document}